\documentclass[a4paper, 10pt]{article}
\usepackage{graphicx}
\newcommand{\beq}{\begin{equation}}
\newcommand{\eeq}{\end{equation}}
\newcommand{\beqa}{\begin{eqnarray}}
\newcommand{\eeqa}{\end{eqnarray}}

\def\opone{\leavevmode\hbox{\small1\normalsize\kern-.33em1}}

\title{Non-realism : deep thought or a soft option ?}
\author{Nicolas Gisin \\
\it \small   Group of Applied Physics, University of Geneva, 1211 Geneva 4,    Switzerland}

\date{\small \today}

\begin{document}
\maketitle

\section{Introduction}\label{intro}
%==================================
In recent years the violation of Bell's inequality has often been interpreted as either a failure of locality or of realism (or of both). The problem with such a claim is that it is not clear what realism in this context should mean. Sometimes realism is defined as the hypothesis that every physical quantity always has a value and that measurements merely reveal these predetermined values. That is, realism is identified with determinism. But if so, then, first, why should one use the word local realism instead of local determinism? And second, Bell's inequality can be stated and proven without any assumption about determinism. Consequently, determinism is not the issue and a meaningful definition of realism has to be found elsewhere.

In order to analyse what realism could mean in the context of Bell inequalities, let us start from the basic assumption underlying them. For simplicity we restrict ourselves to two parties, named as usual today, Alice and Bob. Alice and Bob have a choice between a number of possible measurement settings; each measurement leading to one result among a set of possible results. Actually, since Bell's inequalities are not limited to physics, but can be applied to any kind of multiparty correlations, we prefer to use a slightly more general framework. Hence, let's denote $x$ and $y$ Alice and Bob's inputs (the physicist's measurement settings), respectively, and denote $a$ and $b$ their respective outcomes (the measurement results). A correlation is then the conditional probability distribution $p(a,b|x,y)$ that Alice and Bob's outcomes are $a$ and $b$, given that their inputs are $x$ and $y$, respectively.

So far, so good. However, the situation summarized by the correlation $p(a,b|x,y)$ may actually arise out of a statistical mixture of different situations traditionally labeled by $\lambda_1$ for the first possible situation, $\lambda_2$ for the second one, and so on. These $\lambda$'s may include the usual quantum state $\Psi$; they may also include all the information about the past of both Alice and Bob. Actually, the $\lambda$'s may even include the state of the entire universe, except for the two inputs: $x$ and $y$ should be independent of $\lambda$, i.e. $p(x)=p(x|\lambda)$ and $p(y)=p(y|\lambda)$. But $\lambda$ may as well be much more restricted\footnote{Don't think of $\lambda$ as an old fashion local hidden variable. Think of $\lambda$ as the physical state of the systems as described by any possible future theory. Hence, studying Bell's inequality tells
us something about any possible future theory compatible with today's experimental observations.}: the only constraint on $\lambda$ is that it should not contain any information about the choice of the inputs $x$ and $y$.

Without loss of generality, each conditional correlation can be expanded: $p(a,b|x,y,\lambda)=p(a|x,y,\lambda)\cdot p(b|x,y,a,\lambda)$.
The locality assumption is then that, for any given "state of affair" $\lambda$, what happens on Alice side does not depend on what happens on Bob's side, and vice-versa: $p(a|x,y,\lambda)=p(a|x,\lambda)$ and $p(b|x,y,a,\lambda)=p(b|y,\lambda)$. Consequently, the general assumption underlying all Bell's inequalities reads:
\beq \label{locCondition}
p(a,b|x,y)=\int d\lambda~\rho(\lambda)~ p(a|x,\lambda)\cdot p(b|y,\lambda)
\eeq
where $\rho(\lambda)\ge0$ is the (possibly unknown) probability distribution of the parameter $\lambda$.
Technically, the additional variable lambda is assumed to belong to a set equipped with a measure such that the integral in (\ref{locCondition}) is well defined \cite{Pitowsky}. Another technical reminder is that whenever the set of possible inputs and outcomes are finite, then the set of correlations $\{p(a,b|x,y)\}$ satisfying (\ref{locCondition}) is convex with a finite number of vertices. Each facet of this set corresponds to a Bell inequality. Hence, a correlation $p(a,b|x,y)$ violates a Bell inequality if and only if it can't be decomposed as in (\ref{locCondition}).

Again, the interpretation of (\ref{locCondition}) is as follows: The local probabilities of outcome $a$ on Alice's side and outcome $b$ on Bob's side depends only on the local inputs, $x$ and $y$ on Alice and Bob's sides, respectively, and on the state of affairs, denoted $\lambda$. Hence, given the variable lambda, the probabilities of outcome $a$ on Alice side, for input $x$, $p_A(a|x,\lambda)$, and similarly on Bob's side, $p_B(b|y,\lambda)$, are independent. Accordingly, condition (\ref{locCondition}) is clearly about conditional independence. The motivation for this independence assumption is that since Alice and Bob are spatially separated, all they could do (consciously or oblivious to them) is to exploit some previously established common strategy (described by $\lambda$). This is the locality assumption. But is there also a realism assumption hidden in (\ref{locCondition})?

 Note that condition (\ref{locCondition}) can be formulated within the formalism of standard quantum physics, though we know it is not satisfied by quantum physics:
 \beq
p(a,b|x,y)=Tr(X_a\otimes Y_b ~\rho)\ne \int d\lambda~\rho(\lambda)~ Tr(X_a\lambda_A)\cdot Tr(Y_b\lambda_B)
 \eeq
 where $X_a$ and $Y_b$ are the eigenprojectors of the observables $X$ and $Y$ with eigenvalues $a$ and $b$, respectively. The set of all $\lambda$'s is the set of all pairs of quantum states, $\lambda=(\lambda_A,\lambda_B)$, where the first element of the pair is a valid quantum state $\lambda_A$ of Alice's quantum system and the second element a valid quantum state $\lambda_B$ of Bob's system. This clearly shows that, besides the locality assumption, condition (\ref{locCondition}) doesn't contain any further assumption incompatible with quantum physics.

Many physicists, not familiar with Bell inequalities, get scared when one talks about nonlocality and may thus prefer to write "incompatible with local realism", hoping to avoid nonlocality\footnote{This is even more surprising when one realizes that classical physics has been much more severely nonlocal than quantum physics until the advent of general relativity \cite{incompleteRelativity}; indeed, Newton's law of universal gravitation predicts the possibility of instantaneous signaling across the entire universe by merely moving some masses that, according to Newton's theory, would immediately modify the gravitational field everywhere. Hence, physics provided us with a nonlocal description of Nature during all of its history, except a mere decade, approximately between 1915 and 1925! Note however, that Newton never claimed his theory to be complete \cite{Travis}; it is only his followers that, like e.g. Laplace, elevated Netwon's deterministic equations to some sort of religious belief of ultimate truth.}. From my experience, this is due to a confusion between the kind of nonlocality encountered in quantum physics and the locality condition familiar in relativity \cite{incompleteRelativity}. The fact is that nonlocality does not imply the possibility of signalling, i.e. some nonlocal correlations, for instance those predicted by quantum physics, can't be used to communicate from Alice to Bob, nor from Bob to Alice. Hence no signalling and nonlocality are different concepts, the former is essential for relativity, the later is a well confirmed prediction of quantum physics.

\section{Analysis of possible hiding places for realism}
%=======================================================
Could it be that the abundant usage of the terminology "local realism" is merely a collective bleating of a vacuous phraseology, that is a kind of soft option? or is there something more to it?

Let us look at Fig. 1 that summarizes the situation and let's first concentrate on Alice's side (Bob's side is completely analogous). We see 3 parts: the input (on the top), the outcome (on the bottom) and in-between a black box with blurred boundaries. The details of this black box are unimportant for our discussion, it suffices that it does not extend to Bob's side, i.e. that the two parties can be clearly and unambiguously distinguished. This distinguishability condition can easily be satisfied by a suitably large distance separating Alice and Bob. Let us emphasize that the content of the black box could be anything. It could contain, for instance a bunch of smart physicists equipped with powerful computers or with some quantum particle shelved in some quantum memories, or anything else, classical, quantum or described by some future theory not yet discovered. In particular we don't need to make any realistic assumption about the content of the black box. For example, claiming that the content of the black box is not real doesn't change the situation: the question whether the correlation $p(a,b|x,y)$ can be decomposed as in (\ref{locCondition}) remains unaffected by such a claim.

\begin{figure}[h]
\begin{center}
\includegraphics[width=0.8\linewidth]{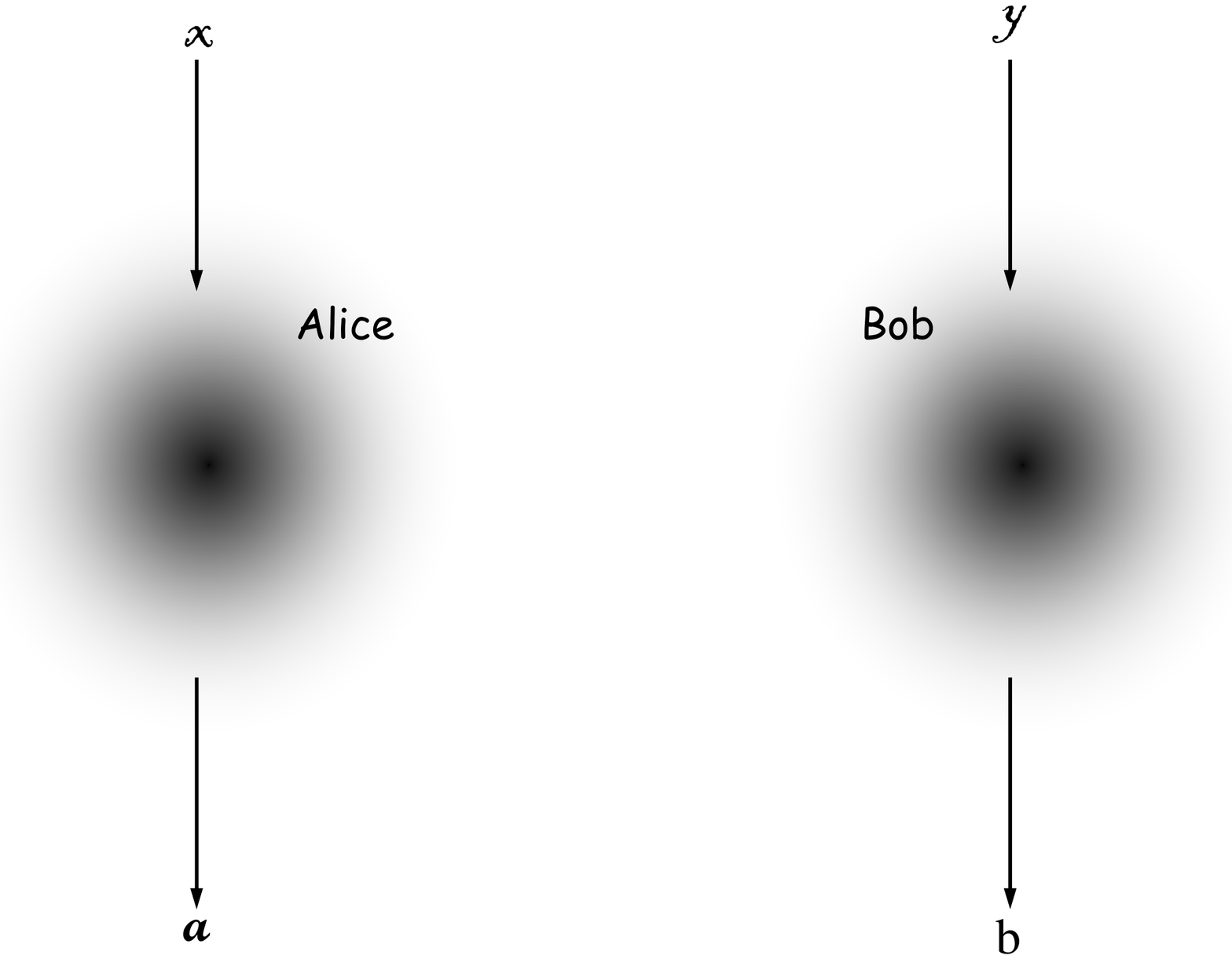}
\end{center}
\caption{\it For each run of the experiment, Alice and Bob each input one value of $x$ and of $y$ into their black boxes; the latter then return one and only one outcome $a$ and $b$ to Alice and Bob, respectively. Note that in order to test condition (\ref{locCondition}) the experiment has to be repeated many times until the statistics allows one to infer a good approximation of the probability $p(a,b|x,y)$.}
\end{figure}

Hence, let's have a closer look at the two other parts of Alice, her input $x$ and outcome $a$. Here we find a possible hint of what realism could mean. Indeed, it is vital for the sake of the argument that both $x$ and $a$ are classical data, i.e. data that can be copied, stored, broadcast, compared and processed by humans. Clearly, if the input $x$ can not be processed by Alice, in particular if Alice can't decide which input she wants to apply for each run of the experiment, or if Alice can't understand the outcome $a$ as one and only one outcome for each run, then we can't make sense of the entire argument depicted in Fig. 1.

It seems we have found one - and only one - place where realism plays a role: the inputs and outcomes have to be real, i.e. the players Alice and Bob should have direct access to their inputs and outcomes. More precisely, they should have the freedom to choose the inputs (or the freedom to choose the device to which they wish to delegate this choice, typically a random number generator of their choice); and they should be able to read the outcomes (in their mind, or, again, to a freely chosen alternative memory device). Note that nothing in the argument is said about the way the outcomes are produced, deterministic or not, under partial influence by the player or totally independent of her/his decisions/actions/wishes.

So, what could non-realism mean? Logically, either that the inputs are not real, or the outcomes (or both). The first alternative would mean that Alice can't freely decide which input to use for each given run of the experiment. Well, this is a logical possibility. But is it an interesting one? I believe no, because if humans are not assumed to have the possibility to decide which experiment to perform, and when to perform them, then there is no way to test any scientific theory: this would be the end of Science. Hence, assuming non-realism of the inputs would amount to a sort of suicide of Science. Let's thus concentrate on the assumption that the outcomes are not really real. What could this mean? Well, here we encounter the infamous quantum measurement problem: possibly measurements don't have one result but lead to a superposition state of the universe corresponding to all possible measurement results. Here I see two ways of developing this argument. First, the usual many-world view. Let me merely state that I consider this view as uninteresting (though logically consistent, like solipsism, another logically consistent though uninteresting view): taken fully seriously, the many world view leaves no space for freedom; indeed, it implies full determinism\footnote{One may think that this is similar to classical physics; however there is a huge difference. In classical physics things are logically separated, hence there is room for a hypothetical mind-body interface (something like Descartes' pineal gland) through which humans (and animals) can freely act upon the material world. Because of entanglement, this escape is impossible in quantum physics. I believe the many world view is incompatible with freedom, hence uninteresting to me. Moreover, I don't see any explanatory power: it merely elevates the linearity of the Schrodinger equation again to some sort of religious belief of ultimate truth.}.

There is, however, a second side to the assumption that the outcomes are not really real: indeed, one could speculate that it takes some time before the black box outputs some definite outcome. If this time is longer than usually assumed, then there could be enough time for a sub-luminal hidden communication between Alice and Bob \cite{Franson}. This raises the following question: when is a quantum measurement finished? Is it as soon as a detector fires, as implicitly assumed by almost all experimentalists? Or should one wait until a human becomes conscious of the result (and here John Bell would ask whether this human should hold a PhD?!?). The general idea that it may take some significant but finite time for a quantum measurement to output a (classical) result is interesting, especially if models can be experimentally tested, see e.g. \cite{Salart08}.

\section{Conclusion} \label{concl}
%=================================
In conclusion, the claim that the observation of a violation of a Bell inequality leads to an alleged alternative between nonlocality and non-realism is annoying because of the vagueness of the second term \cite{Norsen,Laudisa}. The only place where non-realism could make some sense in the context of Bell inequalities is actually the old quantum measurement problem. This is indeed an interesting problem and I am confident that some day physics will have very interesting things to say about it. However, it is not specific to Bell inequalities. Moreover, it is unclear in which sense it could solve the locality problem \cite{Maudlin,Gilder}. Indeed, as long as one can distinguish Alice from Bob, the locality conundrum remains. Hence, all violations of Bell's inequality should be interpreted as a demonstration of nonlocality. Moreover, once nonlocality is accepted new findings become possible, like, e.g. the security of quantum key distribution against any (individual) attack by eavesdroppers only limited by no-signaling \cite{nosignaling}.

Let me finish on a personal note.
For me, quantum nonlocality is an established fact. Yet, it remains mysterious: how does Nature organize its book-keeping to know which measurements should produce nonlocal correlations? Is Nature using an enormously - monstrously - vast Hilbert space to keep track of which physical systems (particles or modes) are entangled? For me this question, like the quantum measurement question, is a real physics question: some day Science will have something meaningful to say about both of these questions. We will then realize that these two questions are essentially different and that the answers open entirely new fields of research for physics.

%\small

\section*{Acknowledgment}
It is a pleasure to thank Stefano Pironio, Rob Thew and Travis Norsen for reading and commenting earlier drafts of this note.
%This work has been supported by the EC under project QAP (contract n. IST-015848) and by the Swiss
%NCCR {\it Quantum Photonics}.

\end{document}